\documentclass[usenatbib]{mn2e}
\usepackage[pdfpagemode={UseOutlines},bookmarks,bookmarksopen,colorlinks,linkcolor={black},citecolor={black},urlcolor={black}]{hyperref}
\usepackage{natbibmnfix}
\usepackage{astrojournals}
\usepackage{epsfig}
\usepackage{amssymb}
\usepackage{multirow}
\usepackage{multicol}
\usepackage{color}
\usepackage[varg]{txfonts}
\newcommand{\msun}{\rm M_\odot}

\title[Black holes in stellar--mass binary systems: expiating original spin?] {Black holes in stellar--mass binary systems: expiating original spin?}

\author[Andrew King \& Chris Nixon]{
Andrew King$^{1,2,3}$ \& Chris Nixon$^{1}$\\
$^{1}$ Theoretical Astrophysics Group, Department of Physics \& Astronomy, University of Leicester, Leicester LE1 7RH, UK\\
$^{2}$ Anton Pannekoek Institute, University of Amsterdam, Science Park 904, 1098 XH Amsterdam, Netherlands\\ 
$^{3}$ Leiden Observatory, Leiden University, Niels Bohrweg 2, NL-2333 CA Leiden, Netherlands  
}

\date{Accepted 2016 June 30. Received 2016 May 27; in original form 2016 May 27}
\pubyear{2016}
\begin{document}
\label{firstpage}
\pagerange{\pageref{firstpage}--\pageref{lastpage}}
\maketitle

\begin{abstract}
We investigate systematically whether accreting black hole systems are likely to reach global alignment of the black hole spin and its accretion disc with the binary plane. In low--mass X--ray binaries (LMXBs) there is only a modest tendency to reach such global alignment, and it is difficult to achieve fully: except for special initial conditions we expect misalignment of the spin and orbital planes by $\sim 1$~radian for most of the LMXB lifetime. The same is expected in high--mass X--ray binaries (HMXBs). A fairly close approach to global alignment is likely in most stellar--mass ultraluminous X--ray binary systems (ULXs) where the companion star fills its Roche lobe and transfers on a thermal timescale to a black hole of lower mass. These systems are unlikely to show orbital eclipses, as their emission cones are close to the hole's spin axis. This offers a potential observational test, as models for ULXs invoking intermediate--mass black holes do predict eclipses for ensembles of $\gtrsim 10$ systems. Recent observational work shows that eclipses are either absent or extremely rare in ULXs, supporting the picture that most ULXs are stellar-mass binaries with companion stars more massive than the accretor.
\end{abstract}

\section{Introduction}
\label{intro}
Accreting black holes frequently have their spins at least initially misaligned from the angular momentum of the mass reservoir feeding them. This is generic for supermassive black holes (SMBH) \citep[cf][]{King:2006aa} and is possible in stellar--mass binary systems, particularly if they have undergone a supernova explosion. But any misalignment must evolve as accretion begins. The differential Lense--Thirring  precession of 
disc orbits produces viscous torques on the accretion disc which try to make everything axisymmetric. In stellar--mass binaries, the flux of mass from the companion with angular momentum parallel to the binary axis usually overwhelms these torques in the outer disc, which stays in the binary plane as a result. But close to the black hole the Lense--Thirring effect generally wins, and the inner disc plane rapidly co-- or counter--  aligns with the spin plane on the local precession time \citep{Scheuer:1996aa,King:2005aa}. The transition between the outer disc, aligned with the binary orbit, and the inner disc, aligned with the hole spin, occurs either in a smooth warp \citep{Bardeen:1975aa} or (for larger misalignments) an abrupt break \citep{Nixon:2012aa,Nixon:2012ad,King:2013aa}.

We call this configuration -- hole spin and inner disc aligned, but both misaligned from the binary orbit -- {\it central alignment}. We expect this kind of alignment for most discs around compact objects because the the Lense--Thirring effect establishes it very quickly in the inner disc after accretion on to the black hole begins, or resumes. This is also likely for SMBH in active galactic nuclei \citep{King:2006aa,King:2007ab}. But if accretion from a binary companion continues for an extended time, the system may also tend towards a state of {\it global alignment}, where spin, disc and orbital rotation are all parallel or (possibly for the spin) antiparallel.

The relative orientation of the hole's spin and the binary axis has a significant effect on the observable properties of accreting stellar--mass black--hole binary systems, so the question of how close a system is to global alignment is important. Studies of it so far either consider individual systems \citep{Martin:2008aa, Maccarone:2002aa,Maccarone:2015aa} or the effect on one method of trying to measure black hole spin \citep{Steiner:2012aa}, which assumes that candidate systems are close to global alignment. Our aim here is to give a systematic picture of whether various types of accreting binaries approach global alignment, including whether this is expected in various models of ultraluminous X--ray sources (ULXs).

\section{torques}
To check whether a given black--hole binary approaches global alignment we assume that the system has already reached central alignment, i.e. of the spin and inner disc planes. We also assume that the inner disc is connected to the outer disc by a smooth warp -- if the inclination is large enough to have caused disc breaking there is little prospect of global alignment, as even central alignment can be disrupted by rapidly--precessing disc rings \citep[disc `breaking' and `tearing':][]{Nixon:2012aa,Nixon:2012ad}. 

The torque between the disc and the black hole trying to bring about global alignment transfers to the hole a fraction $\beta \lesssim 1$ of the Kepler specific angular momentum $j_w = (GMR_w)^{1/2}$ at the characteristic warp radius
\begin{equation}
R_{w} \simeq (a\alpha)^{2/3}\left(\frac{2R}{H}\right)^{4/3}R_g
\label{warp}
\end{equation}
\citep[cf][]{Natarajan:1998aa}, so we write it as
\begin{equation}
{\bf G}_{\rm align} = \beta\dot Mj_w{\bf e}_{\rm orb}
\label{align}
\end{equation}
where ${\bf e}_{\rm orb}$ is the unit vector parallel to the orbital rotation. Here $\dot M$ is the disc accretion rate (strictly, at the warp radius, but usually equal to the mass transfer rate from the companion star), $a$ is the Kerr spin parameter, $\alpha \sim 0.1$ is the standard disc viscosity parameter, $H/R \sim 0.02$ is the local disc aspect ratio, and $R_g = GM/c^2$ is the hole's gravitational radius. From a geometric view, $\beta \sim \sin\theta$, and we shall adopt this for simplicity, so that $\beta$ decreases as alignment proceeds.
Using (\ref{warp}) we find
\begin{equation}
j_w = (a\alpha)^{1/3}\left(\frac{2R}{H}\right)^{2/3}\frac{GM}{c}.
\label{jw}
\end{equation}
There is a second (spinup) torque on the black hole as it gains mass from the innermost stable circular disc orbit (ISCO). This has magnitude 
\begin{equation}
G_{\rm spinup} =\dot M_hj_{\rm isco},
\label{spinup}
\end{equation}
where $\dot M_h \leq \dot M$ is the accretion rate at the black hole (which cannot for example exceed the Eddington value) and  $j_{\rm isco} \sim GM/c < j_w$ is the specific angular momentum at the ISCO. This acts to increase or decrease the black hole angular momentum
\begin{equation}
J_h = \frac{GM^2a}{c}
\label{bham}
\end{equation}
according as accretion is respectively prograde or retrograde.

\section{Global alignment}
We can now check the evolution of the black hole spin vector towards global 
alignment after the companion star has transferred a mass $M_{\rm tr}$ through the accretion disc. The calculation below complements the derivation in \cite{King:2005aa}. This explains the geometry of the alignment process, but does not specify the timescale for it to occur in a given system. In contrast, we give here an estimate of the timescale, independently of the details of warped disc dynamics \citep[see][]{Nixon:2016aa}.

After the transfer of a mass $M_{\rm tr}$, the alignment torque (\ref{align}) has added a component $M_{\rm tr} j_w$ parallel to the orbital rotation by interacting with the spin at the warp radius. We note that this torque is a combination of Lense-Thirring precession and viscous damping \citep{King:2005aa}. Central alignment means that the spinup torque (\ref{spinup}) has simultaneously increased the magnitude of the black--hole spin as
\begin{equation}
J_h = J_{h0} + M_{\rm acc}j_{\rm isco},
\label{acc}
\end{equation}
where $J_{h0}$ was the original value, and $M_{\rm acc} \leq M_{\rm tr}$ is the mass accreted by the black hole. (This may be smaller than $M_{\rm tr}$, as the accretion rate may be super--Eddington for example.) This spinup does not change the original angle $\theta_0$ of ${\bf J}_h$ to the orbital axis, but the alignment torque ${\bf G}_{\rm align}$ does. So after the mass transfer the angle $\theta_f$ of the spin vector to the orbital axis is given by
\begin{equation}
\tan\theta_f \simeq \frac{J_h\sin\theta_0}{J_h{\cos\theta_0} + M_{\rm tr}j_w\sin\theta_0}
\label{final}
\end{equation}
It is straightforward to show (by induction; see Appendix~\ref{app}) that as mass is added iteratively and $\theta$ decreases this equation holds exactly for constant $J_h$, and to first order in ${\rm d}M_{\rm tr}$ if spin magnitude evolution is included. Equation (\ref{final}) can be rearranged to give
\begin{equation}
\frac{M_{\rm tr}}{M} \simeq \frac{0.1a^{2/3}}{(10\alpha)^{1/3}}\left(\frac{50 H}{R}\right)^{2/3}[\cot\theta_f - \cot\theta_0]. 
\label{crit}
\end{equation}
This shows that a modest approach to global alignment ($\theta_f \sim 1$~rad) requires the transfer of a mass 
\begin{equation}
M_{\rm tr} \sim 0.1a^{2/3}M,
\label{modest}
\end{equation} 
but a tighter approach ($\theta_f \sim 0.1$~rad) requires 
\begin{equation}
M_{\rm tr} \sim a^{2/3}M.
\label{tight}
\end{equation}
We see that {\it complete} global alignment ($\theta_f = 0$)  is {\it impossible} for any transferred mass unless $\theta_0 = 0$. The only realistic way of arranging this with $\theta_0 \ne 0$ is for the accretion torque (\ref{spinup}) to spin the hole up from an initially retrograde value $J_{h0} < 0$ through zero.

\begin{figure}
  \center{\includegraphics[width=\columnwidth]{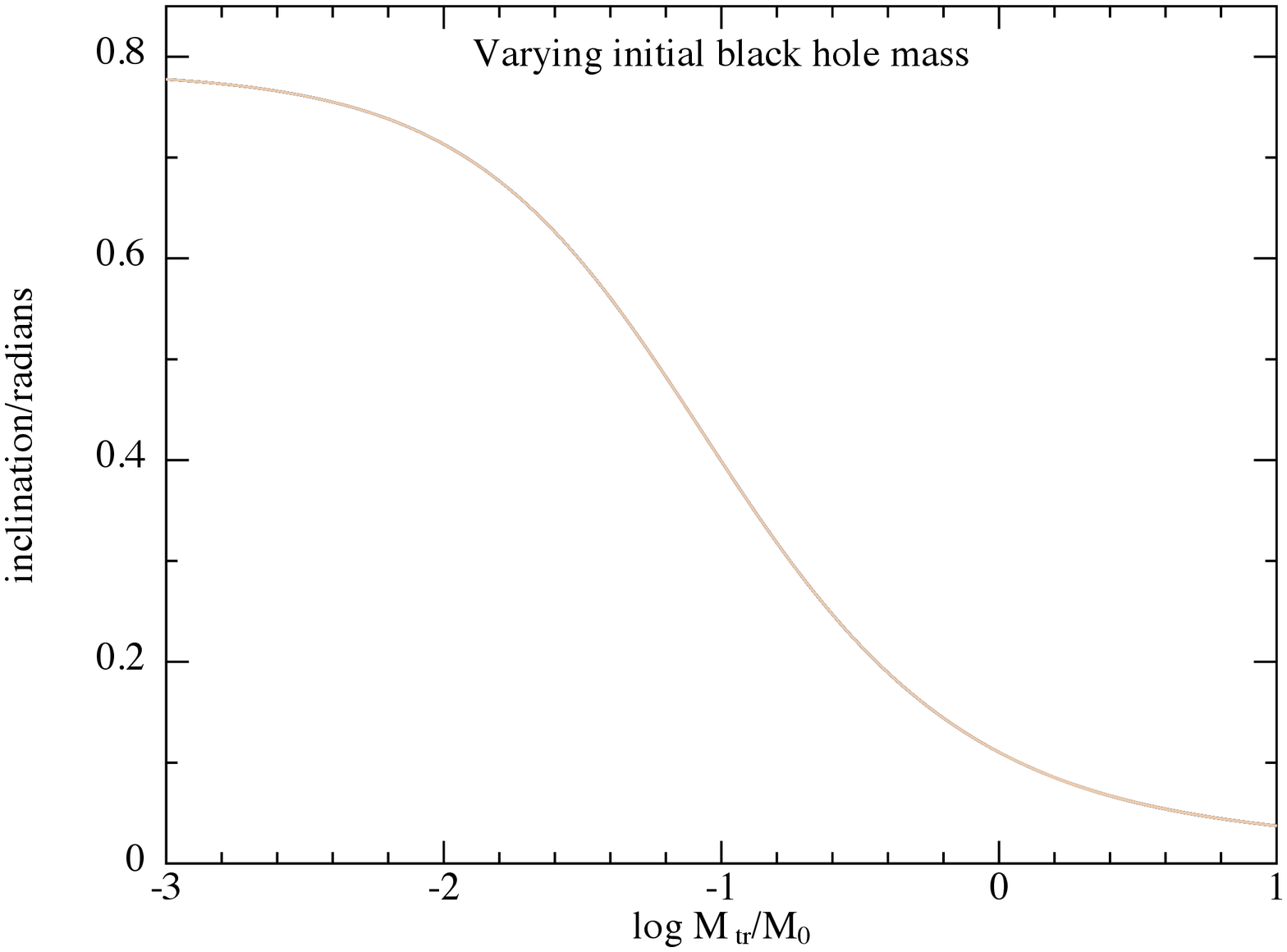}}
  \center{\includegraphics[width=\columnwidth]{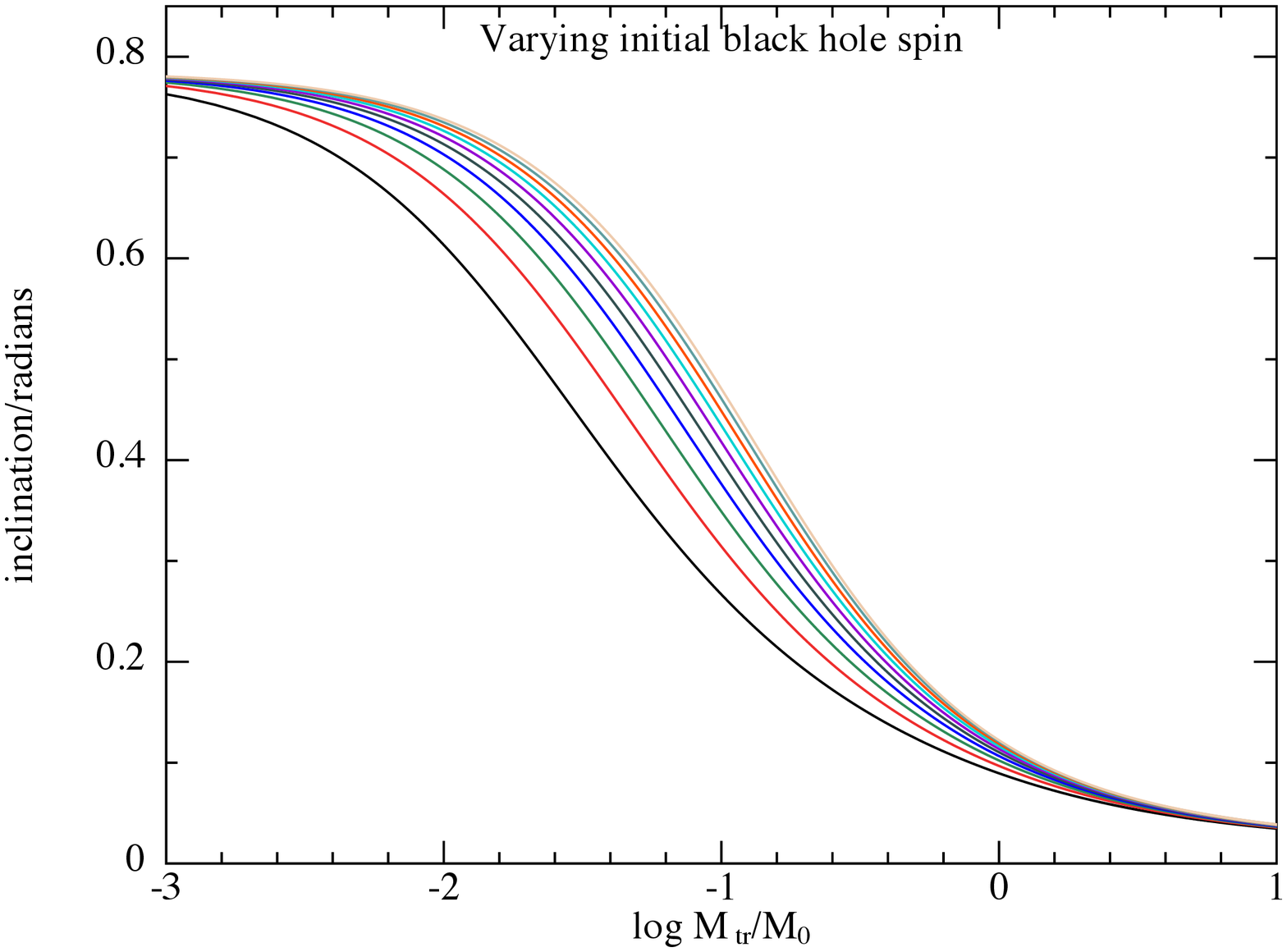}}
  \center{\includegraphics[width=\columnwidth]{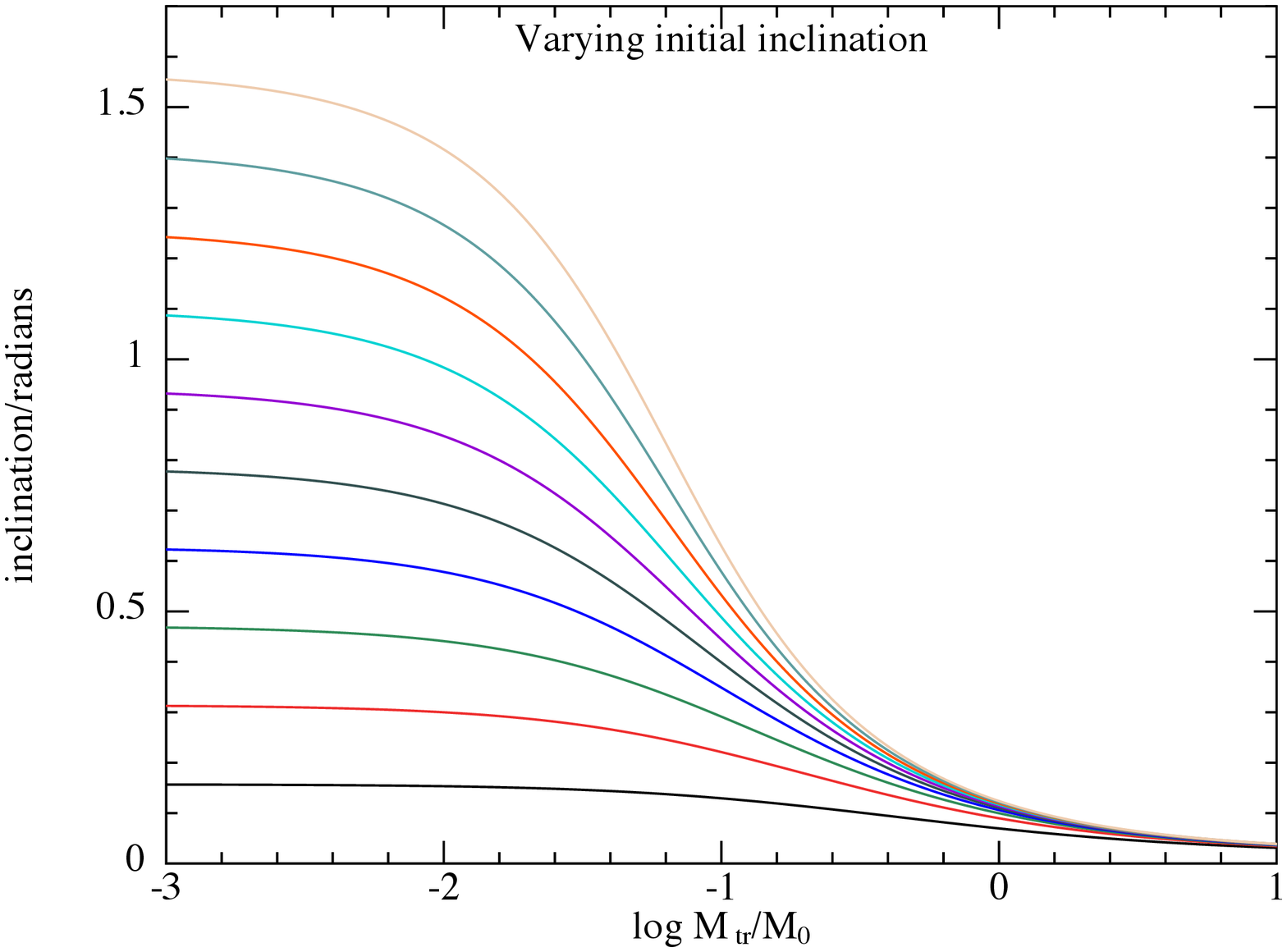}}
  \caption{The three panels describe the evolution of the disc--BH inclination angle as mass is transferred following (\ref{final}). The default parameters are initial BH mass $M_0=5\msun$, inclination $\theta_0 = \pi/4$ and spin $a=0.5$, and we have also assumed $\beta \sim \sin\theta$. Top panel: the curves correspond to varying the BH mass, $M_0$, from $5$ to $15\msun$ in steps of $1\msun$. Middle panel: the curves correspond to varying the initial BH spin from 0.1 to 1 in steps of 0.1. Bottom panel: the curves correspond to varying the initial misalignment angle from $\pi/20$ to $\pi/2$ in steps of $\pi/20$. The top panel shows that for the same value of $M_{\rm tr}/M_0$ the evolution is invariant for different values of $M_0$. In each case the hole's spin does not move significantly until $\sim 0.1M$ ($0.5\msun$) has been transferred. Complete alignment requires a mass $\sim M_0$, as predicted.}
  \label{figs}
\end{figure}    

We show the solutions of these equations for a variety of parameters in Fig.~\ref{figs}. These show the evolution of the misalignment as mass is accreted, transferring misaligned angular momentum to the hole, for (a) different mass black holes, (b) different initial spins and (c) different initial misalignment angles. As predicted by (\ref{tight}) a mass of order $0.1M$ is required to move the spin angle significantly. These basic results (\ref{acc}, \ref{crit}) have straightforward consequences for the various types of black--hole binaries.

\subsection{Standard X--ray binaries}
In low--mass X--ray binaries (LMXBs) the companion mass $M_2$ is small compared  with $M$, so $M_{\rm tr} < M_2 \ll M$. The same result holds for wind--fed high--mass X--ray binaries (HMXBs), as the black hole accretes only a tiny fraction of the mass lost by the companion star  (even though this may have a mass $> M$). So we have $M_{\rm tr} \lesssim \dot M_{\rm Edd}.t_{\rm HMXB} \lesssim 0.1\msun$, where we have taken an Eddington rate $\dot M_{\rm Edd} \lesssim 10^{-7}~{\rm \msun\,yr^{-1}}$ appropriate for a $10\msun$ black hole, and a generous HMXB lifetime of $10^6$~yr. Then (\ref{acc}) shows that $J_h$ remains effectively constant in both HMXBs and LMXBs. Equations (\ref{modest}, \ref{tight}) show that without rather contrived initial conditions only a modest approach to global alignment (i.e. $\theta_f \sim 1$~rad ) is possible in LMXBs, while HMXBs do not move significantly to global alignment at all. Observations of the slightly evolved LMXB GRO 1655--40 also agree with our conclusions, as these show that the spin axis -- revealed by the direction of the jet in this system -- is  far from the binary axis \citep[cf][and references therein]{Martin:2008aa}. We note that this difficulty in reaching full alignment leaves these types of binaries susceptible to disc breaking and tearing, which may explain a variety of the observed properties of LMXBs, including state transitions and QPOs \citep{Nixon:2014aa}.

\subsection{Ultraluminous X--ray Sources}
The HMXB systems considered above naturally evolve to the point where the companion star fills its Roche lobe. Because the companion is generally more massive than the black hole, mass transfer shrinks the binary, and so ultimately proceeds on the thermal timescale of the companion. 
(This also happens in the non-ULX microquasar GRO 1655--40 because the companion is expanding across the Hertzsprung gap -- see \citealt{Martin:2008aa} for a discussion.) This gives very high mass transfer rates, which are strongly super--Eddington for the black hole \citep[cf][]{King:2000aa} and offers a natural model for ultraluminous X--ray sources \citep[cf][]{King:2001aa}. Similar mass transfer rates occur in long--period binaries where the companion is less massive than the black hole, but massive enough for rapid nuclear evolution \citep{Rappaport:2005aa}. In both cases we expect $M_{\rm tr} \sim M$. Then (\ref{modest},\ref{tight}) show that ULXs are likely to be close to global alignment for most of their lifetimes. The spin behaviour is less clear, as in both cases the Eddington limit may mean that the black hole accretes rather little mass and angular momentum during the relatively short (thermal or nuclear timescale) ULX phase. The movement towards global alignment is important, as the ULX property comes from tight geometric collimation of the accretion luminosity around the black hole spin axis \citep[cf][]{King:2001aa,Begelman:2006aa,King:2009aa}. Since the spin moves towards the binary axis, it is unlikely that any ULX of this type would show orbital eclipses. 

The other class of models for ULXs invokes accretion on to an intermediate--mass black hole (IMBH), with mass $M \gtrsim {\rm few }\times 100\msun$, large enough to make the luminosity of the ULX both isotropic and sub--Eddington \citep[cf][]{Colbert:1999aa}. This probably requires rapid nuclear--timescale mass transfer from a fairly massive evolved companion \citep[similar to the picture by][who considered stellar--mass black holes]{Rappaport:2005aa}. For such systems the binary mass ratios are $\sim 0.01 - 0.1$, implying companion Roche--lobe radii $R_2$ which are fractions $\sim 0.1 - 0.2$ of the binary separation $a$, since $R_2/a \simeq 0.462(M_2/M)^{1/3}$. 

A simple geometric argument now shows that an ensemble of more than about 10 such systems should have eclipses in at least one case. For binary inclination $i$ we need $\cos i < R_2/a$  for an eclipse. So the probability of no eclipse in a given case is $1 - R_2/a$, and for $n$ such systems is $(1 - R_2/a)^n \simeq 1 - nR_2/a$. This no--eclipse probability becomes small for sample sizes 
\begin{equation}
n \gtrsim  \frac{a}{R_2} \gtrsim 2\left(\frac{M}{M_2}\right)^{1/3} \sim 10 - 20.
\label{eclipse}
\end{equation}

\section{Conclusions}
We have investigated whether various accreting black hole systems are likely to reach global alignment of the black hole spin and its accretion disc with the binary plane. A fairly close approach to this state is likely in systems where the companion star fills its Roche lobe and transfers mass to a lower--mass black hole. Such systems are promising candidates for ULXs, and are unlikely to show eclipses as their emission cones are close to the hole's spin axis. This offers a potential observational test, as models for ULXs invoking accretion from stellar--mass companions on to intermediate--mass black holes do predict eclipses for an ensemble of $\gtrsim 10$ systems. \cite{Middleton:2016aa} recently showed that eclipses are either absent or extremely rare in among all ULXs for which variability has been measured, in agreement with our result that stellar-mass ULXs should not eclipse because they are close to global alignment.

In standard low--mass X--ray binaries there is a modest tendency to reach global alignment, so except for special initial conditions (such as initially retrograde black hole spin) we would expect a misalignment of the spin and orbital planes $\gtrsim 1$ radian. This agrees with the conclusions of  \cite{Maccarone:2002aa,Maccarone:2015aa}, and weakens those of \cite{Steiner:2012aa}. It increases the systematic error in attempting to measure black hole spin by comparing the area of the event horizon with that expected from the measured black hole mass. Finally, in high--mass X--ray binary systems, neither spinup nor global alignment is likely.

\section*{Acknowledgments}
CN is supported by the Science and Technology Facilities Council (grant number ST/M005917/1). The Theoretical Astrophysics Group at the University of Leicester is supported by an STFC Consolidated Grant. We used {\sc splash} \citep{Price:2007aa} for Fig.~\ref{figs}.

\bibliographystyle{mn2e}
\bibliography{nixon}

%%%%%%%%%%%%%%%%% APPENDICES %%%%%%%%%%%%%%%%%%%%%

\appendix

\section{}
\label{app}

Equation~\ref{final} 
is exact if the spin magnitude evolution is neglected, and valid to first order in ${\rm d}M_{\rm tr}$ if this effect is included. We can show this by induction. Thus
\begin{equation}
  \tan\theta_1 = \frac{J_{\rm h1}\sin\theta_0}{J_{\rm h1}\cos\theta_0 + {\rm d}m j_{\rm w}\sin\theta_0}\,,
\end{equation}
where $J_{\rm h1} = J_{\rm h0} + {\rm d}m j_{\rm isco}$, and
\begin{equation}
  \tan\theta_2 = \frac{J_{\rm h2}\sin\theta_1}{J_{\rm h2}\cos\theta_1 + {\rm d}m j_{\rm w}\sin\theta_1}\,,
\end{equation}
which both follow from (\ref{final}). Now we substitute the equation for $\tan\theta_1$ rearranged as
\begin{equation}
  \cos\theta_1 = \frac{\sin\theta_1\left(J_{\rm h1}\cos\theta_0 + {\rm d}m j_{\rm w}\sin\theta_0\right)}{J_{\rm h1}\sin\theta_0}
\end{equation}
into the equation for $\tan\theta_2$ to get
\begin{equation}
  \tan\theta_2 = \frac{J_{\rm h2}\sin\theta_0}{J_{\rm h2}\cos\theta_0 + 2{\rm d}m j_{\rm w}\sin\theta_0 + {\rm d}m^2 \frac{j_{\rm isco}}{J_{\rm h1}}j_{\rm w}\sin\theta_0}\,,
\end{equation}
which to first order in ${\rm d}m$, or exactly if the spin magnitude evolution is ignored, is as if the equation were evaluated with $M_{\rm tr}=2{\rm d}m$. Neglecting the spin magnitude evolution holds for $M_{\rm tr}j_{\rm isco} \ll J_{\rm h0}$, which implies $M \gg f(a)M_{\rm tr}/a$ (where $f(a)$, the angular momentum of the ISCO in units of $GM/c$, is of order unity). When this requirement is breached, the equation must be solved iteratively with the spin magnitude evolution included.

%%%%%%%%%%%%%%%%%%%%%%%%%%%%%%%%%%%%%%%%%%%%%%%%%%

\bsp
\label{lastpage}
\end{document}